\documentstyle[preprint,aps]{revtex}
\tighten
\def\etap{\eta^\prime}
\def\etam{$\eta$ }
\def\etapm{$\eta^\prime$ }

\begin{document}
\preprint{BIHEP-TH-97-21}
\title{Electromagnetic Transition Form Factor of Pseudoscalar Meson and
$\eta-\etap$ Mixing}
\author{
{\bf Jun Cao$^b$,
Fu-Guang Cao$^{a,c}$, Tao Huang$^{a,b}$, and Bo-Qiang Ma$^{a,b,c}$}\\
\bigskip
$^a$ CCAST (World Laboratory), P.O. Box 8730, Beijing 100080, P.~R.~China\\
$^b$ Institute of High Energy Physics, Academia Sinica,
P.O. Box 918, Beijing 100039, P.~R.~ China\thanks{Mailing address.}\\
$^c$ Institute of Theoretical Physics, Academia Sinica,
P.O. Box 2735, Beijing 100080, P.~R.~China
}
\date{\today}
\maketitle

\begin{abstract}
The electromagnetic transition form factors 
of $\eta$ and $\etap$ 
are calculated in the light-cone perturbation theory. 
We show that it is unreliable 
to determine
the \etam-\etapm mixing angle without any additional 
normalization conditions other than
their decay widthes to two photons. 
The possible intrinsic $c\overline{c}$
component in the flavor singlet is investigated. 
The heavy quark pair has
distinct properties from the light ones in electromagnetic transition
processes of pseudoscalar mesons. It is possible to explore the size of
$c\overline{c}$ component and our numerical results disfavor a large portion of
$c\overline{c}$ component.


\end{abstract}

\section{Introduction}
\parindent 20pt
The recent CLEO
experiments related with \etam and \etapm supply possibilities to extract
the information about the structures of 
these two pseudoscalar mesons. For example, the CLEO
collaboration \cite{cleo1,cleo2} has reported very large branching ratios
for inclusive \etapm production
\begin{equation}
B( B^{\pm}\rightarrow\etap X_s;2.0 
{\rm GeV}\le P_{\etap}\le 2.7{\rm GeV}) =
( 7.5\pm1.5\pm1.1)\times 10^{-4},
\end{equation}
and for the exclusive decay $B^{\pm}\to\etap{K}^{\pm}$
\begin{equation}
B(B^{\pm}\rightarrow\etap{K}^{\pm}) =
(7.8^{+2.7}_{-2.2}\pm1.0)\times10^{-5} .
\end{equation}
To explain the abnormally large production of \etapm in the standard model,
either large portion of intrinsic  
$|c\bar{c}\rangle$ component in \etapm \cite{hal,cheng}
or large coupling of \etapm to $gg$ \cite{gg}, even both of them, must be
concluded. At the same time, the data of the $\pi\gamma$, $\eta\gamma$ and
$\etap\gamma$ transition form factors at higher energies reported by CLEO
\cite{cleo3} suggest that \etapm may have very different
non-perturbative properties from $\pi^0$ and \etam, while the latter two
have similar wave functions. 
It is reasonable because the physical \etam and
\etapm states 
consist dominantly of
flavor SU(3) octet $\eta_8$
and singlet $\eta_0$, respectively. The usual mixing scheme reads
\begin{mathletters}
\label{mixing}
\begin{eqnarray}
|\eta\rangle &=&\cos(\theta)|\eta_8\rangle-\sin(\theta)|\eta_0\rangle\ ,\\
|\etap\rangle&=&\sin(\theta)|\eta_8\rangle+\cos(\theta)|\eta_0\rangle\ .
\end{eqnarray}
\end{mathletters}
The mixing angle can be evaluated in various ways but a standard procedure
involves the diagonalization of the $\eta$ and $\eta^{\prime}$ mass matrix, 
which
at the lowest order in the chiral perturbation theory yields a mixing angle
$\theta\approx-10^{\circ}$. The inclusion of ${\cal O}(p^4$) corrections
\cite{mix} to
the relation results in significant changes, which yields 
$\theta\cong-20^{\circ}$.
\par
At the same time, the $\eta\gamma$ and $\etap\gamma$ transition form
factors have received 
extensive theoretical attentions recently. Jakob {\it et
al} \cite{kroll} extracted the mixing angle from the $\eta\gamma$ and
$\etap\gamma$ trantion form factors in the modified hard scattering approach
\cite{li}, which takes into account the transverse degrees of freedom as well
as the Sudakov form factor, 
and got $\theta=-18^\circ\pm2^\circ$ with a Gaussian
wave function in transverse part. In their approach the chiral anomaly for
Goldstone boson is used to determine the decay constants of pseudoscalar
mesons, including the \etapm meson. 
The parameters related to transverse degree
of freedom for all three mesons, $\pi$, \etam and \etapm, are assumed to be
identical for simplicity. With the same approach, 
a more general mixing scheme
involving two mixing angles was investigated \cite{feldmann} recently. 
The charm decay constant of the $\eta^\prime$ meson was estimated to 
be within the range of -65 MeV $\leq f_{\eta^\prime}^c \leq$ 15 MeV. 
Choi and Ji \cite{ji} 
showed their form factor
prediction  for both $\theta=-10^\circ$ and $-23^\circ$
by using a simple relativistic constituent quark model with a
Gaussian wave function motivated from light-cone quantization and
the Brodsky-Huang-Lepage (BHL) prescription, which connects the oscillator
wave function in the center of mass frame
with the light-cone wave function \cite{bhl}.
They fixed the
Gaussian parameter by radiative decay of pseudoscalar and vector
mesons and
found that they are in good agreement with experimental data up to a rather
large $Q^2$. Anisovich {\it et al} \cite{ani} studied the meson-photon
transition form factor by assuming a nontrivial hadron-like $q\overline{q}$
structure of the photon in the soft region. The data on the $\pi^0\gamma$
form factor are used to fix the soft photon wave function. Assuming
the universality of pseudoscalar meson wave function 
in the ground-state they found
that the $\eta\gamma$ and $\etap\gamma$ transition form factors and
the branching ratios for $\eta\to\gamma\gamma$ and
$\etap\to\gamma\gamma$ are in perfect agreement with the data.
\par
It is interesting that an analysis \cite{caofg}    
performed in the standard light-cone formalism
\cite{2157,Brodsky,BrodskyPQCD} 
shows that there is still a gap between the
theoretical calculations and the data at currently accessible energies.
The conclusion is reasonable since only the lowest Fock state 
and the lowest order
contributions (without radiative correction) are taken into account. To
fit the data, some efforts must be made to account for the higher Fock
state and higher order contributions. The soft photon wave function adopted
in Ref.~\cite{ani}, which is extracted from the $\pi^0\gamma$ data, 
may be considered as
a good phenomenological description for higher order contributions
presenting at the soft photon vertex. The soft photon may behave much like a 
hadron
whose wave function evolves by exchanging gluons between the quark-antiquark
pair. On the other hand, the normalization and the parameters of the 
pionic wave
function are fixed when only the valence Fock state is taken into
account in the light-cone Fock state expansion. 
The obtained expression is for
the valence Fock state. If we go beyond the light-cone Fock state
expansion by giving up the constraint from $\pi\to\mu\nu$, to which only the
$|q\overline{q}\rangle$ component contributes, and adjust the
parameters
in the wave function in a reasonable region, 
a perfect agreement with the data can be obtained.
The same parameter fits the data of the $\eta\gamma$ and $\etap\gamma$ 
transition form factors, too.
A light-cone constituent quark model calculation \cite{ji} also justified
that some of the higher Fock state contributions 
in the light-cone perturbation
theory may be expressed by a light-cone constituent quark model with only
$q\overline{q}$ being taken into account. 
\par
With the aids of models 
accounting for the higher Fock state and higher order
contributions, informations about the wave functions of the $\eta$
and $\etap$ mesons can be extracted from the data of the $\eta\gamma$
and $\etap\gamma$ transition from factors. 
In section II, 
a brief review on the $\pi\gamma$ transition will be given to explain some
subtle aspects, including the Melosh rotation 
connecting the light-cone wave
function and the ordinary equal-time wave function, 
the parameter fixing and the possible corrections. In Sec. III,
analyses on the $\eta\gamma$ and $\etap\gamma$ transitions will be presented. 
By introducing a
$SU_f(3)$ broken wave function, it is found that the theoretical predictions
fit
the data very well. At the same time, our results are not sensitive to the
mixing angle because similar mixing patterns present 
in both the form factors and the
$\eta(\etap)\to\gamma\gamma$ amplitudes, 
while the latter are used as normalization
conditions. Without additional 
normalization conditions being used, determining the
mixing angle by fitting the data of 
the $\eta\gamma$ and $\etap\gamma$ transition form
factors is not reliable. 
The flavor singlet $|\eta_0\rangle$ may have large portions of 
$|c\overline{c}\rangle$ and $|gg\rangle$ components.
In Sec. IV we discuss the possible intrinsic $|c\overline{c}\rangle$
contributions, which have distinct $Q^2$ behavior from the light quark
contributions. Our results disfavor a large portion of $|c\overline{c}\rangle$
component in the $\etap$ wave function expected in Refs.~\cite{hal,cheng}
and such a conclusion is in agreement with a number of recent
investigations from other viewpoints \cite{sic1,sic2,sic3,sic4}.
The last section contains the conclusions and summary.

\section{The $\pi\gamma$ transition form factor}
\indent\par
The approach adopted here for the $\eta\gamma$ and
$\eta^\prime\gamma$ transition form factors will be very similar to that of
$\pi\gamma$ \cite{caofg,2157} 
A brief review of the $\pi\gamma$ transition form 
factor will
help to reveal some subtle aspects in this approach.

\par
An analysis on the $\pi\gamma$ transition in 
the light-cone perturbation theory,
based on light-cone quantization and the light-cone Fock state expansion, was
first investigated by Brodsky and Lepage \cite{2157}. Factorization into a soft
part (the pionic wave function $\psi(x,k_i)$) and a hard part (the hard
scattering amplitude $T_H$) is verified since soft interactions between
initial and final quarks in $T_H$ all cancel due to the fact 
that hadron states are
color singlets. Thus the form factor can be written in a factorization form
as
\begin{equation}
\label{form}
F_{\pi\gamma}(Q^2)=2\sqrt{n_c}(e_u^2-e_d^2)\int^1_0d x\int
\frac{d^2k_\perp}{16\pi^3} \psi(x,k_\perp) T_H(x,k_\perp,Q) \ ,
\end{equation}
where the hard scattering amplitude $T_H$ is given by 
\begin{equation}\label{hard}
T_H(x,k_\perp,Q)=\frac{q_\perp\cdot(x_2q_\perp+k_\perp)}
{q_\perp^2(x_2q_\perp+k_\perp)^2} +(1\leftrightarrow2) \ ,
\end{equation}
with $x_2=1-x$ and $q_\perp^2=Q^2$.
When $Q^2$ is large, the quark transverse momentum $k_\perp$ can be omitted
in $T_H$ compared with $q_\perp$. The integral over $k_\perp$ results in a
distribution amplitude (DA), $\phi(x,Q)
=\int\frac{d^2 k_{\perp}}{16 \pi^3}\psi(x,k_{\perp})$, 
which evolves to the asymptotic form,
$\sqrt{3}f_\pi x(1-x)$, as $Q^2\to\infty$.
Thus we get the asymptotic prediction for  
the $\pi\gamma$ transition form factor,
$F_{\pi\gamma}(Q^2\to\infty)=\frac{2f_\pi}{Q^2}$.
However, the transverse momentum is not negligible at currently accessible
energies
to be compared with the experimental data \cite{caofg}. 
Taking into account the Melosh rotation \cite{melosh,cckp},
which connects the
equal-time spin and the light-cone spin, the light-cone wave function of
a pseudoscalar meson can be written as \cite{hms,MaZ,cchm}
\begin{equation}\label{wavem}
\Psi_R(x,k_\perp)=\frac{1}{\sqrt{2(m^2+k^2)}}\left[
m\left(\chi_1^\uparrow\chi_2^\downarrow-\chi_1^\downarrow\chi_2^\uparrow
\right)-(k_1+i k_2)\chi_1^\downarrow\chi_2^\downarrow-(k_1-i
k_2)\chi_1^\uparrow\chi_2^\uparrow\right]\psi(x,k_\perp)\ ,
\end{equation}
where the index 1 (2) means the quark (antiquark) and $\uparrow$ ($\downarrow$)
is the light-cone helicity.
The Melosh rotation is one of the most 
important ingredients of the light-cone
quark model and it can be applied to explain
the ``proton spin puzzle" \cite{puzzle}.
The spacial wave function $\psi(x,k_\perp)$ is modeled as \cite{bhl}
\begin{equation}
\label{bhlw}
\psi(x,k_\perp) = A \exp\left[-b^2 \frac{k_\perp^2+m^2}{x(1-x)}\right]\ 
\end{equation}
with the Brodsky-Huang-Lepage prescription which 
connects the oscillator wave function 
in the equal-time
frame with that in the light-cone frame. Such an exponential ansatz for the wave 
function
have a simple form, direct physical explanation (oscillator form in center
of mass frame), and good end-point behaviors.
From the exponential form (\ref{bhlw}) follows the
distribution amplitude
\begin{equation}\label{eq8}
\phi(x)=x(1-x)\frac{A}{16\pi^2b^2}\exp\left[-b^2\frac{m^2}{x(1-x)}\right],
\end{equation}
which is very close to the asymptotic form \cite{hms}. 
It is well known that the DA evolves very slowly. For a DA close to 
the asymptotic form, such as eq.~(\ref{eq8}), the evolution makes little
difference. Furthermore, it is difficult to take into account the evolution
while the transverse momentum, and thus the wave function but not DA,
is involved. We will just neglect the evolution in the following 
calculations for simplicity. 
Many recent analyses on pionic
non-perturbative properties \cite{kroll,hms,pionwave,rady} favor 
the asymptotic form of DA 
rather than the Chernyak-Zhitnitsky (CZ) 
form \cite{cz}.
Therefore, the BHL model for the wave function is favored by
fitting experimental data.
\par
In the pionic case two important constraints have been derived \cite{bhl}
from the $\pi\to\mu\nu$ and $\pi\to\gamma\gamma$ decay amplitudes:
\begin{eqnarray}
\int^1_0 d x \int\frac{d^2k_\perp}{16\pi^3} \psi(x,k_\perp) &=&
\frac{f_\pi}{2\sqrt{3}} \ ,\label{cstr1}\\
\int^1_0 d x \psi(x,k_\perp=0) &=&\frac{\sqrt{3}}{f_\pi} \ .\label{cstr2}
\end{eqnarray}

\par
Firstly, it is notable that $m$ 
in the non-perturbative
wave function, i.e., Eqs. (\ref{wavem}) and (\ref{bhlw}), 
is the
constituent quark mass while contributions from higher helicity states
($\lambda=\lambda_1+\lambda_2=\pm1$) 
will be proportional to 
the current quark mass in the corresponding hard scattering
amplitude and thus can be
ignored due to the fact that helicity must be flipped at one vertex.
However, it is
not the case for the heavy quark whose current quark mass is almost the same as
the constituent quark mass. Secondly, 
the gauge invariance requires that 
the $|q\overline{q}\rangle$ Fock state in pion contributes exactly one half
to the full decay amplitude of $\pi\to\gamma\gamma$ \cite{bhl}.
Therefore, higher
Fock states must have contributed the other half to $F_{\pi\gamma}$ as
$Q^2\to0$. 
While $Q^2$ increases, the contributions from higher Fock states
reduce. 
It is not strange that the numerical results in Ref.~\cite{caofg}
are below the data. 
At currently accessible energy scale, the
$|q\overline{q}\rangle$ component contributes about $80\sim90 \%$ to the
$\pi\gamma$ transition form factor. Thirdly, the one-loop radiative
correction was calculated in Ref.~\cite{oneloop}. If the asymptotic 
distribution amplitude is
used, the size of the one-loop correction 
is independent of the factorization
scale $\mu$ and less than 15\% for $Q^2>3$ GeV. Otherwise, different
non-perturbative wave functions result in different sizes of corrections. An
appropriate choice of scale $\mu$ may reorganize the expansion series to
reduce the one-loop correction. Numerical analyses \cite{rady} show that
$\mu=Q$ provides a good choice of the factorization scale. It is
accompanied by small one-loop corrections even for a broad
distribution amplitude of CZ type.
Since the correction is small, we will not go beyond the lowest order
calculation to avoid such a complexity in the following. Finally,
taking into account the Melosh rotation in the pionic wave function
changes the
light-cone perturbation prediction only for a small amount, because the
constraint (\ref{cstr1}) should be changed as
\begin{equation}
\int^1_0 d x
\int\frac{d^2k_\perp}{16\pi^3}\psi(x,k_\perp)\sqrt{\frac{m^2}{m^2+k^2_\perp}}=
\frac{f_\pi}{2\sqrt{3}} \ ,\label{cstr3}
\end{equation}
and constraint (\ref{cstr2}) remains unchanged.
As a consequence, the parameters in the wave function should be different. The
numerical results are shown in Fig.~1 with parameters
$b^2=0.848$ GeV$^{-2}$, $A=32.7 $ GeV$^{-1}$
from the constraints (\ref{cstr1})  and (\ref{cstr2}) (the dashed line) 
and $b^2=0.414$ GeV$^{-2}$, $A=25.6 $ GeV$^{-1}$
from the constraints (\ref{cstr3})  and (\ref{cstr2}) (the solid line). 
The quark mass $m$ does
not affect a lot and we adopt $m=300$ MeV here. 

\section{The $\eta\gamma$ and $\etap\gamma$ form factors in SU$_f$(3)}
\indent\par
The transition form factor at zero momentum transfer is connected with the
two-photon decay width by 
\begin{equation}
F_{R\gamma}(0)=\sqrt{\frac{4}{\alpha^2\pi
M_R^3}\Gamma(R\to\gamma\gamma)} \ ,\label{eta0}
\end{equation}
where $R$ represents a pseudoscalar meson.
 From the axial anomaly in the chiral limit of QCD, we have 
\begin{equation}
\lim_{Q^2\to 0} F_{\gamma R}(Q^2)=\frac{1}{4\pi^2f_R}
\end{equation}
for $\pi^0$ and $\eta$. This prediction may not hold for $\eta^\prime$ due
to the larger mass of s-quark. In addition, it might be broken because
$\eta^\prime$ is an unlikely candidate for the Goldstone boson. We will not
relate the wave function with decay constant in this work.
To  be consistent with the experimental analysis, we use the same values
adopted in Ref.~\cite{cleo3}:
\begin{eqnarray*}
\Gamma(\pi\to\gamma\gamma)  =&7.74\ {\rm eV}\ ,
~~~~~~~~~~~~ & f_\pi=92.3\ {\rm MeV}\ ;\\
\Gamma(\eta\to\gamma\gamma) =& 0.463\ {\rm keV}\ ,   
~~~~~~~~~~~~ & f_\eta=97.5\ {\rm MeV}\ ;\\
\Gamma(\etap\to\gamma\gamma)  =&4.3\ {\rm keV}\ .
~~~~~~~~~~~~ & ~
\end{eqnarray*}

\par
Unlike the pion decay, only the constraint (\ref{cstr2}) is available to
normalize the amplitude for the $\eta\gamma$ and $\etap\gamma$
transitions, i.e.
to determine two paremeters in Eq. (\ref{waves}) with the mixing angle
$\theta$ as an input. The CLEO collaboration reported their pole fit results
as $\Lambda_\pi=776\pm10\pm12\pm16$ MeV, $\Lambda_\eta=774\pm11\pm16\pm22$
MeV, and $\Lambda_{\eta^\prime}=859\pm9\pm18\pm20$ MeV. The L3 collaboration
\cite{l3}
also presented their pole mass, $\Lambda_{\eta^\prime}=900\pm46\pm22$ MeV. It
suggests that the non-perturbative properties of $\pi$ and \etam are very
similar. It is also consistent with the physical intuition since both $\pi$
and \etam are in SU$_f$(3) octet and are pseudo-Goldstone particles. It
is a natural choice to set $b_8=b_\pi$. The singlet should have
different properties. But for simplicity, we will let $b_0=b_\pi$ at first. 
Different choices for $b_0$ will be discussed, too.
\par
As the case of $\pi\gamma$, the light-cone predictions for the 
$\eta\gamma$ and $\etap\gamma$ transition form factors are
smaller than the data while only the lowest Fock state 
and the lowest order contributions are taken into account.
To compare them with the data, we assume that higher Fock state and higher 
order contributions have
similar fraction sizes in the transition form factors 
of  all three
pseudoscalar particles. These contributions can be estimated from
the $\pi\gamma$ form factor and 
included into the $\eta\gamma$ and $\etap\gamma$
form factors. The pole form 
\begin{equation}
F^{pole}_{\pi\gamma}(Q^2)=\frac{F_{\pi\gamma}(0)}{1+Q^2/\Lambda_\pi^2} 
\end{equation}
is used as the experimental value. The $\eta\gamma$ and $\etap\gamma$ form
factors, after this correction, are obtained as
\begin{equation}\label{corrected}
F_{R\gamma}(Q^2)=F^{LC}_{R\gamma}(Q^2) \frac{F^{pole}_{\pi\gamma}(Q^2)}
{F^{LC}_{\pi\gamma}(Q^2)},
\end{equation}
where $F^{LC}_{R\gamma}(Q^2)$ is the theoretical prediction in 
the light-cone
calculation with only the lowest Fock state 
and the lowest order contributions being taken into
account.
\par
For certain circumstances, SU$_f$(3) is not a good symmetry due to the
large value of the $s$ quark mass. 
The SU$_f$(3) broken form of wave functions for
flavor singlet $\eta_0$ and octet $\eta_8$ can be modeled as
\begin{mathletters}\label{waves}
\begin{eqnarray}
\eta_0&=&A_0\frac{1}{\sqrt{3}}\left
[\exp\left(-b^2_0\frac{m_q^2+k_\perp^2}{x_1x_2}\right) 
\left(u\overline{u}+d\overline{d}\right)+
\exp\left(-b^2_0\frac{m_s^2+k_\perp^2}{x_1x_2}\right) s\overline{s}\right]
\ ,\label{waver1}\\
\eta_8&=&A_8\frac{1}{\sqrt{6}}\left
[\exp\left(-b^2_8\frac{m_q^2+k_\perp^2}{x_1x_2}\right) 
\left(u\overline{u}+d\overline{d}\right)-2
\exp\left(-b^2_8\frac{m_s^2+k_\perp^2}{x_1x_2}\right) s\overline{s}\right]
\label{waver8}
\end{eqnarray}
\end{mathletters}
from the BHL model.
\par
The transition form factor of SU(3) singlet or octet is
\begin{equation}\label{lcsu3}
F^{LC}_{i\gamma}(Q^2)=2\sqrt{2n_c}\sum_{q=u,d,s}C^q_i\int^1_0d x\int
\frac{d^2k_\perp}{16\pi^3} \sqrt{\frac{m_q^2}{m_q^2+k^2_\perp}}
\psi^q_i(x,k_\perp) T^q_H(x,k_\perp,Q) \ ,
\end{equation}
where $i=0$ or 8 presents the singlet or octet. The charge factors 
are
$C^u_8=e_u^2/\sqrt{6}$,
$C^d_8=e_d^2/\sqrt{6}$,
$C^s_8=-2e_s^2/\sqrt{6}$,
$C^u_0=e_u^2/\sqrt{3}$,
$C^d_0=e_d^2/\sqrt{3}$, and
$C^s_0=e_s^2/\sqrt{3}$.
The hard scattering
amplitude $T^q_H$ is the same as Eq.~(\ref{hard})
and $\psi^q_i(x,k_\perp)$ is
the BHL wave function, Eq.~(\ref{bhlw}),
for each flavor quark with the corresponding $A_i$, $m_q$
and $b_i$ as input parameters. 
The quark mass $m_q$ in the
factor from the Melosh rotation will be adopted as $m_{u,d}=300$ MeV
or $m_s=450$ MeV, depending on the flavor in $\psi^q_i(x,k_\perp)$.
\par
Summing over all flavors, We obtain
the transition form factors of physical $\eta$ and $\etap$ states
from Eq.~(\ref{form})
\begin{mathletters}\label{lcphy}
\begin{eqnarray}
F^{LC}_{\eta\gamma}(Q^2)&=&\cos(\theta)F^{LC}_{8\gamma}(Q^2)-
\sin(\theta)F^{LC}_{0\gamma}(Q^2)\ ,\\
F^{LC}_{\etap\gamma}(Q^2)&=&\sin(\theta)F^{LC}_{8\gamma}(Q^2)+
\cos(\theta)F^{LC}_{0\gamma}(Q^2)\ ,
\end{eqnarray}\end{mathletters}
which, to ensure the feasibility of perturbation theory, should be valid
only for large $Q^2$. 
While fitting the data, we will choose only the data
points with $Q^2\geq2.94$ GeV$^2$.
It is interesting to note that the form factors 
$F^{LC}_{\eta\gamma}(Q^2)$ and 
$F^{LC}_{\etap\gamma}(Q^2)$ follow
the same pattern of mixing as the state, Eq.~(\ref{mixing}).
\par
As $Q^2\to 0$, the gauge invariance requires that Eq.~(\ref{lcphy})
contributes exact one half to $F_{R\gamma}(0)$.  Inputing a mixing angle
$\theta$, $A_1$ and $A_8$ can be fixed by $F_{R\gamma}(0)$. The theoretical
predictions obtained from Eq.~(\ref{corrected}) are shown in Fig.~2. The best
fit value of $\theta$ is $-24^\circ$. However, we find that the differences
between different choices of $\theta$ are so small that it is in fact
unreliable
to determine the mixing angle, e.g. $\theta=-14^\circ$ can not be
excluded by the data.
The reason is that $F_{R\gamma}(0)$, which we use as the normalization 
condition,
shares the same mixing mechanism as the form factor $F_{R\gamma}(Q^2)$. The
only difference comes from the different contributions between the $s$ 
quark and
the $u$($d$) quark at different $Q^2$. 
If $m_s=m_{u(d)}$, the mixing will 
be
meaningless in this approach. The same conclusion also holds as
$Q^2\to\infty$, while all quarks can be treated as massless.
Since the perturbation calculations are only valid for large $Q^2$, this
minor difference is negligible. At present, neither the data nor the
theoretical approach has reached the accuracy to distinguish the $s$ or
$u$($d$) quark in the transition form factors of pseudoscalar mesons. Therefore,
unless other normalization conditions (e.g. $D_s\to\eta l \nu/\etap l \nu$,
$\etap\to\omega\gamma$ and $\omega\to\eta\gamma$, and so on.) are involved, 
it is unreliable 
to determine the mixing angle from the data of the transition
form factors. Inclusion of other normalization conditions brings additional
uncertainties and goes beyond the present work. 

\section{The intrinsic $c\overline{c}$ component}
\indent\par
The ``intrinsic quark" is one of the novel properties
for hadrons and  there have been many examples where
the non-valence ``intrinsic quark" components seem
important \cite{iq}.
The flavor singlet meson $\eta_0$ may have a large  portion of
$c\overline{c}$ component
and strong coupling to gluons through QCD axial anomaly. 
Using operator
product expansion and QCD low energy theorems, the non-perturbative
intrinsic charm content of the \etapm meson was evaluated semi-quantitatively
in Ref.~\cite{hal} to be $f_{\etap}^c=50-180$ MeV, which suffices to explain
the large \etapm production in $B$ decay reported by CLEO. Furthermore,
Ref.~\cite{cheng} suggested $f_{\etap}^c\sim -50$ MeV to
account for the data if $c\overline{c}$ pair is in color octet. Especially,
the sign of $f_{\etap}^c$ was fixed as minus in their work. A similar size,
$f_{\etap}^c=40$ MeV, was argued in Ref.~\cite{chao}, too.
\par
The gluonic
components of $\eta_0$ play no role in electromagnetic transition form
factors because the coupling of two photons to two gluons is very small.
They are invisible here. However, the
$c\overline{c}$ pair has distinct behavior from the light quark pairs.
Unlike the $s\overline{s}$ pair, different sizes of
$c\overline{c}$ component change the form factor a lot. 
In the following, the current
mass and constituent mass of $c$ quark are treated as the same and we adopt
$m_c=1.5$ GeV. The hard scattering amplitude for the heavy quark pair is 
\begin{equation}\label{chard1}
T_H(x,k_\perp,Q)=\frac{q_\perp\cdot(x_2q_\perp+k_\perp)}
{q_\perp^2((x_2q_\perp+k_\perp)^2+m_c^2)} +(1\leftrightarrow2) \ .
\end{equation}
It is noted that the helicity-flip amplitude 
can not be ignored 
since it is proportional to
current quark mass. 
A direct calculation gives the higher helicity contributions as
\begin{equation}\label{flip}
F_{c}^{\lambda=\pm1}(Q^2)=2\sqrt{2n_c}e_c^2\int^1_0d x\int
\frac{d^2k_\perp}{16\pi^3} \psi_c(x,k_\perp) 
\frac{1}{\sqrt{m_c^2+k^2_\perp}}\left(\frac{m_c q\cdot{k}}
{q_\perp^2((x_2q_\perp+k_\perp)^2+m_c^2)} +(1\leftrightarrow2)\right)\ .
\end{equation}
The total contributions of all helicity components can be expressed as
Eq.~(\ref{lcsu3}) by changing the hard scattering amplitude as
\begin{equation}\label{chard}
T_H(x,k_\perp,Q)=\frac{q_\perp\cdot(x_2q_\perp+2k_\perp)}
{q_\perp^2((x_2q_\perp+k_\perp)^2+m_c^2)} +(1\leftrightarrow2) \ .
\end{equation}
The spacial wave function is written as
\begin{equation}\label{wavec}
\psi_c(x,k_\perp) = A_c\exp
\left[-b_0^2 \frac{k_\perp^2+m_c^2}{x(1-x)}\right]\ .
\end{equation}
The $c\overline{c}$ contributions to the transition form factor have very
different $Q^2$ behavior from light quark ones because of the higher
helicity contributions, the heavy quark propagator, and the different wave
function. Comparison with light quark contributions is shown in Fig.~3
with their relative sizes at zero momentum transfer normalized to unity. This
difference may result in observable effects in 
the transition form factors.

Now we explore the form factor with the interested value 
$f^c_{\eta^\prime}=-50$ MeV. 
The decay constant is connected with the wave function as
the pionic case \cite{2157} as
\begin{equation}\label{dc}
f_R^q=\langle0|\overline{q}\gamma^+(1-\gamma^5)q|R\rangle
=2\sqrt{2n_c}\int_0^1 d x \int\frac{d^2k}{16\pi^3}
\psi^{q\overline{q}}(x,k_\perp)\ .
\end{equation}
The above decay constant is multiplied by 
a factor of $\sqrt{2}$ comparing with
the previous sections to be consistent with Refs.~\cite{hal,cheng,chao}.

\par
$A_c$ is easily obtained by the definition of the decay constant, 
Eq.~(\ref{dc}),
with the input $f^c_{\etap}$ and the mixing angle $\theta$. 
The mixing angle is fixed
at $\theta=-20^\circ$ because it is, by our numerical results, still not 
sensitive even when $c\overline{c}$ is included.
At the same time, we neglect the difference between $s$ quarks and 
$u$($d$) quarks in order to concentrate on $c$ quarks, 
i.e., let $m_s=m_{u(d)}=300$ MeV. This approximation
is valid since the difference is minor, which can be seen in last section. 
The obtained form factor, comparing with $f^c_{\etap}=0$, is shown in
Fig.~4.
The corresponding decay constants are
\begin{eqnarray*}
f^{u(d)}_{\etap}=56.8\ {\rm MeV}\ ,&
~~~~~f^{s}_{\etap}=105.4\ {\rm MeV}\ ,&
~~~~~f^{c}_{\etap}=-50.0\ {\rm MeV}\ ;\\
f^{u(d)}_{\eta}=71.1\ {\rm MeV}\ ,&
~~~~~f^{s}_{\eta}=-62.5\ {\rm MeV}\ ,&
~~~~~f^{c}_{\eta}=-18.2\ {\rm MeV}\ .
\end{eqnarray*}
These values for light quarks are similar to that with $f^c_{\etap}=0$
because the $c\overline{c}$ contributions at zero momentum transfer are very
small and change the normalization condition only slightly. 
They are not far from
the naive expectation in SU$_f$(3) limit at $\theta=-20^\circ$:
\begin{eqnarray}
f^u_{\etap}\sim f^d_{\etap}\sim f^s_{\etap}/2
\sim f_\pi/\sqrt{6}=54\ {\rm MeV}\ ,\\
f^u_{\eta}\sim f^d_{\eta}\sim -f^s_{\eta}\sim f_\pi/\sqrt{3}=77\ {\rm MeV}\ .
\end{eqnarray}

\par
Since the flavor singlet has different non-perturbative properties from 
that of the octet, 
the parameter $b_0$ is not necessarily the same as $b_8$. In
fact, with $f^c_{\eta^\prime}=-50$ MeV, 
a very good fit can be obtained by adjusting 
$b_0^2$ to 0.3 GeV$^{-2}$ rather than
$b_0^2=b_\pi^2=0.414$ GeV$^{-2}$. 
However, the decay constants will increase apparently
for light quarks (about 30\% for $u$ and $d$ quark):
\begin{eqnarray*}
f^{u(d)}_{\etap}=74.4\ {\rm MeV}\ ,&
~~~~~f^{s}_{\etap}=119.1\ {\rm MeV}\ ,&
~~~~~f^{c}_{\etap}=-50.0\ {\rm MeV}\ ;\\
f^{u(d)}_{\eta}=73.3\ {\rm MeV}\ ,&
~~~~~f^{s}_{\eta}=-48.8\ {\rm MeV}\ ,&
~~~~~f^{c}_{\eta}=-18.2\ {\rm MeV}\ .
\end{eqnarray*}
This difference will result in apparent increasement 
to the radiative decay of mesons, such as $\rho(\omega,\phi)\to\eta\gamma$,
$\etap\to\rho(\omega)\gamma$ and $\phi\to\etap\gamma$. Choi and Ji \cite{ji}
have found that the same value $b_\pi$ for all of these flavor nonet mesons 
can produce satisfied decay widths for these decay channels. 
An increasement of 30\%,
if $b_0^2=0.3$ GeV$^{-2}$
rather than $b_0=b_\pi$, may exceed too much to fit the data of their 
radiative decays. 
Therefore the above numerical results
disfavor such a choice
and thus disfavor a large portion of $c\overline{c}$ component
as $f^{c}_{\etap}=-50\ {\rm MeV}$. 
\par 
There have been recently a number of investigations 
\cite{sic1,sic2,sic3,sic4} which support
a small $f^{c}_{\etap}$ than that was estimated in Ref.~\cite{hal}. 
Our above conclusion is in agreement with these results
from other considerations. For example, in Ref.~\cite{sic4},
$f^{c}_{\etap}=-12.3 \sim -18.4 \ {\rm MeV}$ was suggested.
To check the consistency of such a possibility in our
approach, we adjust the input to 
$f^{c}_{\etap}=-15 \ {\rm MeV}$ and present the result
in Fig.~4. 
The decay constants for different flavor quarks are
\begin{eqnarray*}
f^{u(d)}_{\etap}=56.0\ {\rm MeV}\ ,&
~~~~~f^{s}_{\etap}=104.6\ {\rm MeV}\ ,&
~~~~~f^{c}_{\etap}=-15.0 \ {\rm MeV}\ ;\\
f^{u(d)}_{\eta}=70.8\ {\rm MeV}\ ,&
~~~~~f^{s}_{\eta}=-62.7\ {\rm MeV}\ ,&
~~~~~f^{c}_{\eta}=-5.5\ {\rm MeV}\ ,
\end{eqnarray*}
with $b_0^2=b_\pi^2=0.414$ GeV$^{-2}$. 
We notice that the obtained results are very close to the
case of $f^{c}_{\etap}=0$ and 
our analysis allow a small 
$f^{c}_{\etap}$ around $-15 \ {\rm MeV}$. 
This is consistent with the results in 
Refs.~\cite{sic1,sic2,sic3,sic4}.

\section{Conclusions and summary}
\indent\par
The electromagnetic transition form factors of $\eta$ and $\eta^\prime$ are
presented comparing with $\pi$ in this paper. 
The numerical results show that there still
exists a gap between the data and the light-cone perturbation
calculation with the Brodsky-Huang-Lepage wave function as
the input for the non-perturbative aspects of the mesons. 
To model the higher Fock state and higher order contributions,
we assume that the 
fraction
size of these contributions are similar to all 
three mesons: pion,
$\eta$, and $\eta^\prime$. The ratio of the data 
of the $\pi\gamma$ transition form
factor to the theoretical calculation is multiplied to the light-cone results
of the $\eta\gamma$ and $\eta^\prime\gamma$ form factors. 
With such a correction, the obtained transition form factors
can be compared with the data. In SU$_f$(3), it is in fact 
unreliable to
determine the mixing angle at present unless additional normalization
conditions other than the $\eta(\eta^\prime)\to\gamma\gamma$ decay widths are
included. The reason is that the same mixing mechanism occurs  in both the
transition form factors and the normalization conditions. 
Both the experimental (especially for $\etap$) and theoretical 
accuracies are not high enough to distinguish the $u$($d$) and $s$ quark 
pairs in these pseudoscalar mesons at high $Q^2$.
\par
The heavy quark pair has different $Q^2$ behavior from the light ones. It is
possible to explore the size of the $c\overline{c}$ component in the flavor
singlet. Our results disfavor a large portion of $c\overline{c}$ component
in $\etap$. 
Such a conclusion is in agreement with a number of recent 
investigations with a small portion of intrinsic charm 
in \etapm from other considerations.


\newpage
\section*{Figure caption}
\begin{description}
\item{Fig. 1.}
The pion-photon transition form factor with different parameters.
The dashed line is the form factor without Melosh rotation. 
The solid line is with Melosh rotation. 
The dashed-dotted line is the
result as $Q^2\rightarrow\infty$,
while the evolution of the BHL wave function is neglected.
\item{Fig. 2.}
The transition form factors with different mixing angles.
\item{Fig. 3.}
The ratio of the $c\bar c$ contribution to the $q\bar q$ ($q$ being light quark)
contribution to the transition form factor,
which is normalized to unity at zero momentum transfer.
\item{Fig. 4.}
The transition form factors with $f_{\eta^\prime}^c=0$ MeV,
$f_{\eta^\prime}^c=-15$ MeV
and $f_{\eta^\prime}^c=-50$ MeV.
\end{description}


\begin{references}

\bibitem{cleo1}
B. Behrens, talk presented at {\it The Second International Conference on B
Physics and CP violation}, March 24-27, 1997, Honolulu, Hawaii.
\bibitem{cleo2}
F. W\"urthwein, hep-ex/9706010.
\bibitem{hal}
I. Halperin, A. Zhitnitsky, Phys. Rev. D {\bf 56}, 7247 (1997). 
For further development, see, 
E.V. Shuryak and A. Zhitnitsky, Phys. Rev. {\bf D 57}, 2001 (1998).
\bibitem{cheng}
H.Y. Cheng and B. Tseng, Phys. Lett. {\bf B415}, 263 (1997).
\bibitem{gg}
D. Atwood and A. Soni, Phys. Lett. {\bf B405}, 150 (1997); 
W.S. Hou and B. Tseng, Phys. Rev. Lett. {\bf 80}, 434 (1998).
\bibitem{cleo3}
The CLEO collaboration, J.~Gronbery {\it et al.},
Phys. Rev. {\bf D 57}, 33 (1998).
\bibitem{mix}
E.P. Venugopal, Barry R. Holstein, hep-ph/9710382.
\bibitem{kroll}
R. Jakob, P. Kroll, and M. Raulfs, J. Phys. G {\bf 22}, 45 (1996).
\bibitem{li}
H.N. Li and G. Sterman, Nucl. Phys. {\bf B381}, 129 (1992).
\bibitem{feldmann}
T. Feldmann and P. Kroll, hep-ph/9711231.
\bibitem{ji}
H.-M. Choi and C.-R. Ji, Nucl. Phys. {\bf A618}, 291 (1997);
Phys. Rev. {\bf D 56}, 6010 (1997);
\bibitem{bhl}
S.J. Brodsky, T. Huang, G.P. Lepage, 
in {\it Particle and Fields 2}, edited by
A.Z. Capri and A.N. Kamal 
(Plenum Publishing Corporation, New York, 1983), p. 143;
T. Huang, {\it Proceedings of XX-th International Conference on High
Energy Physics}, Madison, Wisconsin, 1980, 
edited by L. Durand and L.G. Pondrom,
AIP Con. Proc. No. 69 (AIP, New York, 1981), p. 1000.
\bibitem{ani}
V.V. Anisovich, D.I. Melikhov, and V.A. Nikonov, Phys. Rev. D {\bf 55},
2918 (1997).
\bibitem{caofg}
F.-G. Cao, T. Huang, and B.-Q. Ma, Phys. Rev. D {\bf 53}, 6582 (1996).
\bibitem{2157}
G.P. Lepage and S.J. Brodsky, Phys. Rev. D {\bf 22}, 2157 (1980);
{\bf 24}, 1808 (1981).
\bibitem{Brodsky}
	S. J. Brodsky and G. P. Lepage,
	Phys. Rev. Lett. {\bf 53}, 545 (1979);
	Phys. Lett. {\bf 87B}, 359 (1979).
\bibitem{BrodskyPQCD}
S. J. Brodsky and G. P. Lepage, 
{\it Perturbative Quantum  Chromodynamics}, edited by
A. H. Mueller (Singapore, World Scientific, 1989), p. 93.
\bibitem{sic1}
A. Ali and C. Greub, Phys. Rev. {\bf D 57}, 2996 (1998);
A. Ali, J. Chay, C. Greub, and P. Ko, hep-ph/9712372.
\bibitem{sic2}
T. Feldmann, P. Kroll, and B. Stech, hep-ph/9802409.
\bibitem{sic3}
A.A. Petrov, hep-ph/9712497.
\bibitem{sic4}
F. Araki, M. Musakhanov, and H. Toki,
hep-ph/9803356.
\bibitem{melosh}
E. Wigner, Ann. Math. {\bf 40}, 149 (1939);  
H.J. Melosh, Phys. Rev. D {\bf 9}, 1095 (1974).
\bibitem{cckp}
P.L. Chung, F. Coester, B.D. Keister, and W.N. Polyzou, 
Phys. Rev. C {\bf 37}, 2000 (1988).
\bibitem{hms}
T. Huang, B.-Q. Ma, and Q.-X. Shen, Phys. Rev. D {\bf 49}, 1490 (1994).
\bibitem{MaZ}
B.-Q. Ma, Z. Phys. {\bf A 345}, 321 (1993);
B.-Q. Ma and T. Huang, J. Phys. {\bf G 21}, 765 (1995).
\bibitem{cchm}
F.-G. Cao, J. Cao, T. Huang, and B.-Q. Ma, Phys. Rev. D {\bf 55}, 7107 (1997).
\bibitem{puzzle}
B.-Q.~Ma, J. Phys. {\bf G 17}, L53 (1991); 
B.-Q.~Ma and Q.-R.~Zhang, Z.~Phys. {\bf C 58}, 479 (1993); 
S.J.~Brodsky and F.~Schlumpf, Phys. Lett. {\bf B 329}, 111 (1994); 
B.-Q.~Ma, Phys.~Lett.~{\bf B 375}, 320 (1996);
\bibitem{pionwave}
P. Kroll and M. Raulfs, Phys. Lett. {\bf B387}, 848 (1996);
S. Ong, Phys. Rev. D {\bf 52}, 3111 (1995);
A.V. Radyushkin and R.T. Ruskov, Phys. Lett. {\bf B374}, 173 (1996);
V. Braun and I. Halperin, Phys. Lett. {\bf B328}, 457 (1994);
A.P. Bakulev and S.V. Mikhailov, Mod. Phys. Lett. {A11}, 1611 (1996).
\bibitem{rady}
I.V. Musatov, A.V. Radyushkin, Phys. Rev. D {\bf 56}, 2713 (1997).
\bibitem{cz}
V.L. Chernyak and A.R. Zhitnitsky, Phys. Rep. {\bf 112}, 173 (1984).
\bibitem{oneloop}
F. Del Aguila and M.K. Chase, Nucl. Phys. {\bf B193}, 517 (1981);
E. Braaten, Phys. Rev. D {\bf 28}, 24 (1983);
E.P. Kadantseva, S.V. Mikhailov and A.V. Radyushkin, Sov. J. Nucl. Phys.
{\bf 44}, 326 (1986).
\bibitem{l3}
The L3 Collaboration, CERN-PPE/97-110.
\bibitem{iq}
See, e.g., 
S.~J.~Brodsky, P.~Hoyer, C.~Peterson, and N.~Sakai, Phys. Lett.
{\bf B 93}, 451 (1980);
S.J.~Brodsky and B.-Q.~Ma, Phys. Lett. {\bf B 381}, 317 (1996); 
S.J.~Brodsky and M.~Karliner, Phys. Rev. Lett. {\bf 78}, 4682 (1997);
and references therein. 
\bibitem{chao}
F. Yuan and K.T. Chao, Phys. Rev. D {\bf 56}, 2495 (1997). 
\end{references}
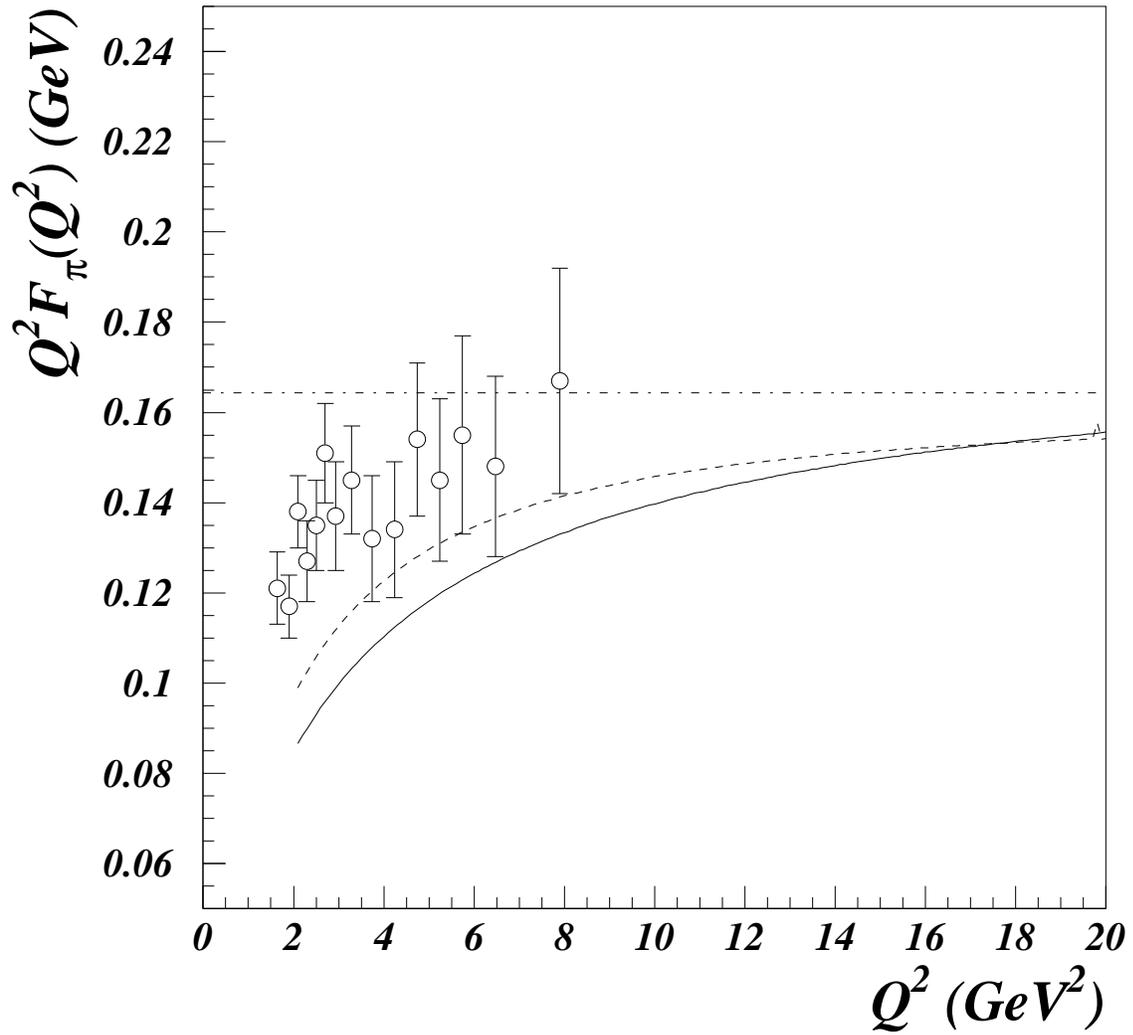
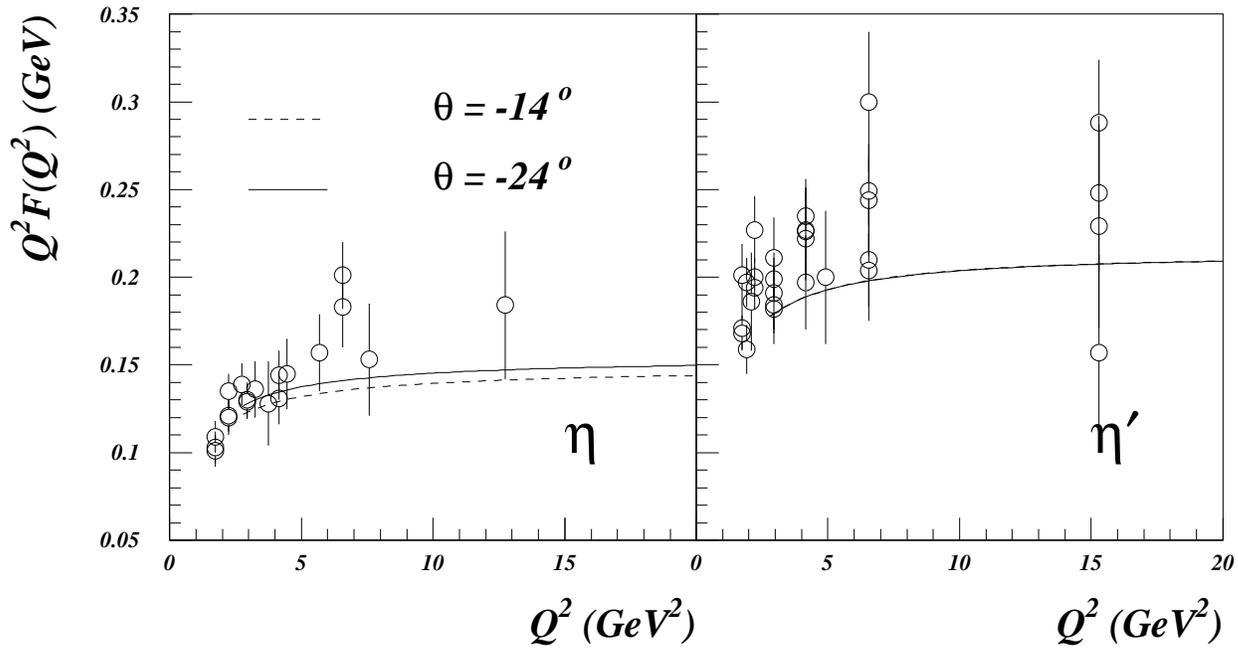
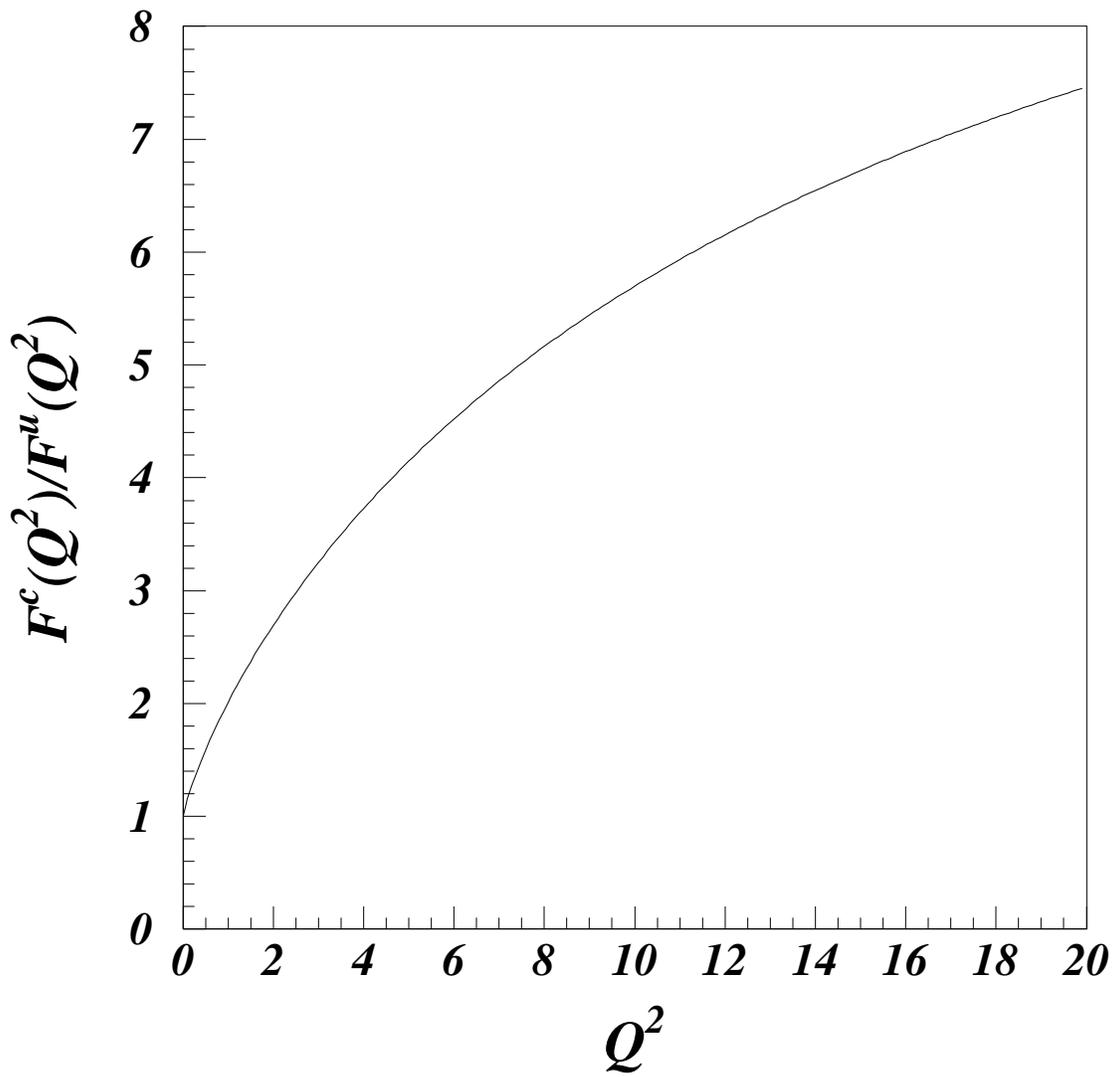
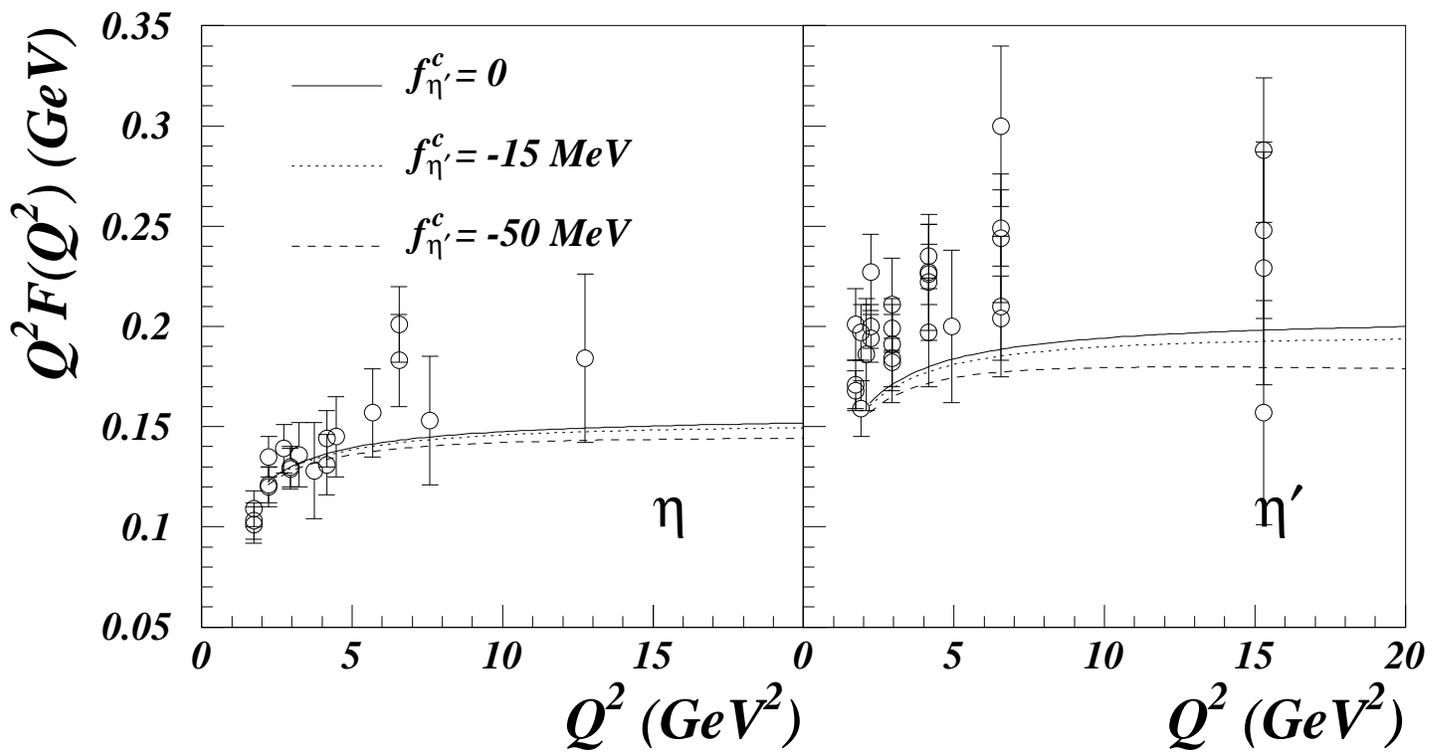
\end{document}